\documentstyle[amssymb]{article}
\newcommand{\Lie}{{{{\frak g}}}} % for Lie algebra
\def\greaterthansquiggle{\raise.3ex\hbox{$>$\kern-.75em\lower1ex\hbox{$\sim$}}}
\def\lessthansquiggle{\raise.3ex\hbox{$<$\kern-.75em\lower1ex\hbox{$\sim$}}}
\newcommand{\beq}{\begin{equation}}\newcommand{\be}{\begin{equation}}
\newcommand{\beqn}{$$}
\newcommand{\eeqn}{$$}
\newcommand{\lam}{{\sigma}}
\newcommand{\eeq}{\end{equation}}
\newcommand{\beqa}{\begin{eqnarray}}
\newcommand{\eeqa}{\end{eqnarray}}
\newcommand{\beqan}{\begin{eqnarray*}}
\newcommand{\eeqan}{\end{eqnarray*}}
\newcommand{\ba}{\begin{array}}
\newcommand{\ea}{\end{array}}

\newcommand{\Sext}{\Sigma_{\mathrm{ext}}}

\newcommand{\Ra}{\Rightarrow}

\newcommand{\Om}{{\Omega}}

\def\nz{\ifmmode {I\hskip -3pt N} \else {\hbox {$I\hskip -3pt N$}}\fi}
\def\zz{\ifmmode {Z\hskip -4.8pt Z} \else
       {\hbox {$Z\hskip -4.8pt Z$}}\fi}
\def\qz{\ifmmode {Q\hskip -5.0pt\vrule height6.0pt depth 0pt
       \hskip 6pt} \else {\hbox
       {$Q\hskip -5.0pt\vrule height6.0pt depth 0pt\hskip 6pt$}}\fi}
\def\rz{\ifmmode {I\hskip -3pt R} \else {\hbox {$I\hskip -3pt R$}}\fi}
\def\cz{\ifmmode {C\hskip -4.8pt\vrule height5.8pt\hskip 6.3pt} \else
       {\hbox {$C\hskip -4.8pt\vrule height5.8pt\hskip 6.3pt$}}\fi}
\def\au{{\setbox0=\hbox{\lower1.36775ex%
\hbox{''}\kern-.05em}\dp0=.36775ex\hskip0pt\box0}}
\def\ao{{}\kern-.10em\hbox{``}}
\newtheorem{Theorem} {Theorem} [section]

\newtheorem{Proposition} [Theorem] {Proposition}

          \newcommand{\N}{{\bf N}}
          \newcommand{\R}{{\bf R}}
{\catcode `\@=11 \global\let\AddToReset=\@addtoreset}
\AddToReset{equation}{section}

\newcommand{\subjclass}[1]{}

\newcommand{\proof}{{\sc Proof:}\ }
\def\scri{\hbox{${\cal J}$\kern -.645em {\raise
      .57ex\hbox{$\scriptscriptstyle (\ $}}}}

\newcommand{\calo}{{\cal O}}

 \newcommand{\eq}[1]{(\ref{#1})}

\newcommand{\commentout}[1]{}

\newcommand{\ee}{\end{equation}} \newcommand{\bea}{\begin{eqnarray}}
\newcommand{\eea}{\end{eqnarray}}
\newcommand{\beaa}{\begin{eqnarray*}}
\newcommand{\eeaa}{\end{eqnarray*}} 

\newcommand{\ra}{\rightarrow}

\begin{document}

\title{The isometry groups of asymptotically flat, asymptotically
empty space--times with timelike ADM four--momentum} 
\author{Robert Beig\thanks{{\em E--mail}: Beig@Pap.UniVie.AC.AT
\protect\newline\indent 
$^\dagger$ On leave of absence from the Institute of
    Mathematics, Polish Academy of Sciences, Warsaw.
  Supported in part by KBN grant \# 2P30209506
 and by the Federal Ministry of Science and Research, Austria.
  {\em E--mail}:    Chrusciel@Univ-Tours.fr} 
  \\ Institut f\"ur Theoretische Physik\\ Universit\"at Wien\\ A--1090
  Wien, Austria\\ \\ Piotr T.\ Chru\'sciel$^\dagger$%\thanks{ 
\\ D\'epartement de Math\'ematiques\\Facult\'e
des Sciences\\ Parc de Grandmont\\ F37200 Tours, France}

\maketitle

\begin{abstract} 
  We give a complete classification of all connected isometry groups,
  together with their actions in the asymptotic region,
  in  asymptotically flat, asymptotically vacuum space--times with
  timelike ADM four--momentum.
\end{abstract}

\section{Introduction}
\label{introduction}

In any physical theory a privileged role is played by those solutions of the
dynamical equations which exhibit symmetry properties. For example,
according to a current paradigm,
there should exist a large class of
isolated gravitating systems  which are expected to settle down
towards a stationary state, asymptotically in time, outside of black
hole regions. If that is the  
case, a classification of all such stationary states would give exhaustive
information about the large--time dynamical behavior of the solutions
under consideration. More generally, one would like to understand the
global structure of all appropriately regular space--times exhibiting
symmetries. Now the local structure of space--times with Killing
vectors is essentially understood, the reader is referred to the book
\cite{Exactsolutions}, a significant part of which is devoted to that
question. However, in that reference, as well as in most works devoted
to those problems, the global issues arising in this context are not
taken into account. In this paper we wish to address the question, what
is the structure of the connected component of the identity of the
group of isometries of space--times which are asymptotically flat in
space--like directions, when the condition of time--likeness of the ADM
four--momentum $p^\mu$ is imposed. Recall that the time--likeness of
$p^\mu$ can be established when the Einstein tensor satisfies a
positivity condition, and when the space--time contains an
appropriately regular spacelike surface, see \cite{ChBeig1} for a
recent discussion and a list of references. Thus the condition of
time--likeness of $p^\mu$ is a rather weak form of imposing global
restrictions on the space--time under consideration. The reader should
note that we do not require  $p^0$ to be positive, so that our
results also apply  to space--times with negative mass, as long as the
total four--momentum is time--like.

In asymptotically flat space--times one expects Killing vectors to
``asymptotically look like'' their counterparts in Minkowski
space--time -- in \cite[Proposition 2.1]{ChBeig1} we have shown that
 {\em at the leading order} this is indeed the case (see also
Proposition \ref{PD.1} below). This  
allows one to classify the Killing vectors into ``boosts'',
``translations'', etc., according to their leading asymptotic
behavior. There exists a large literature concerning the case in which
one of the Killing vectors 
is a time--like translation -- {\em  e.g.},  the
 theory of uniqueness of black holes -- but no exhaustive analysis of
 what Killing vectors are kinematically allowed has been done so
 far. This might be due to the fact, that for Killing vector fields
with a  rotation--type leading order behaviour, the next to
leading order terms are essential to analyse the structure of the
orbits, and it seems difficult to control those without some overly
restrictive hypotheses on the asymptotic behaviour of the metric. In
this work we overcome  this difficulty, and prove the following (the 
reader is referred to Section \ref{proofs} for the definition of a
boost--type domain, and for a detailed presentation of the asymptotic
conditions used in this paper): 

\begin{Theorem}
  \label{T0}
  Let $(M,g_{\mu\nu})$ be a  space--time containing an asymptotically flat
  boost--type domain
  $\Omega$, with time--like (non--vanishing)  ADM four momentum
  $p^\mu$, with fall--off exponent $1/2 < \alpha <1$ and differentiability
  index $k\ge 3$ (see eq.\ \eq{D1.new} below). We shall also assume that
  the hypersurface $\{t=0\}\subset\Omega$ can be 
  Lorentz transformed 
  to a hypersurface in $\Omega$ which is asymptotically orthogonal to $p^\mu$.
  Suppose moreover
  that the Einstein tensor $G_{\mu\nu}$   of $g_{\mu\nu}$ satisfies in
  $\Omega$ the fall--off condition 
\begin{equation}\label{Efalloff}
G_{\mu\nu}= \calo(r^{-3-\epsilon}), \qquad \epsilon > 0\ .
\end{equation}
Let $X^\mu$ be a non--trivial Killing vector field on $\Omega$, let
$\phi_s[X]$ denote its (perhaps only locally defined) flow. Replacing
$X^\mu$ by an appropriately chosen multiple thereof if necessary, one has:
\begin{enumerate}
\item There exists $R_1\ge 0$ such that $\phi_s[X](p)$ is defined for
  all $p\in \Sigma_{R_1}\equiv\{(0,\vec x)\in \Omega:r(\vec x )\ge R_1\}$
and for all $s\in [0,1]$. 
\item There exists a constant $a\in\R$ such that, in local coordinates
on $\Omega$, for all $x^\mu=(0,\vec x)$ as in point 1 we have   
$$
\phi^\mu_1[X]= x^\mu + a p^\mu + \calo_{k}(r^{-\alpha})\ .
$$
\item If $a=0$, then $\phi_1[X](p)=p$ for all $p$ for which
$\phi_1[X](p)$ is defined.
\end{enumerate}
\end{Theorem}

The reader should notice that Theorem \ref{T0} excludes boost-type
Killing vectors. This feature is specific
to asymptotic flatness at spatial infinity, see \cite{BicakSchmidt}
for a large class of vacuum space--times with boost symmetries which
are asymptotically flat in 
light--like directions. The theorem is sharp, in the sense that the
result is not true if $p^\mu$ is allowed to vanish or to be
non--time--like.

When considering asymptotically flat space--times with more than one
Killing vector, it is customary to assume that there exists a linear
combination of Killing vectors the orbits  of which are periodic (and
has an axis --- see below).
However no justification of this property of Killing orbits has been
given so far, except perhaps in some special situations. Theorem
\ref{T0} allows us to show that this is necessarily the case. While
this property, appropriately understood, can be established without
making the hypothesis of 
completeness of the orbits of the Killing vector fields, the
statements become somewhat awkward. For the sake of simplicity let
us therefore assume that we have an action of a connected non--trivial
group $G_0$ on $(M,g_{\mu\nu})$ by isometries. Using Theorem \ref{T0}
together with the results of \cite{ChBeig1} we
can classify all the groups and actions. Before doing that we need to
introduce some terminology. Consider a space--time $(M,g_{\mu\nu})$
with a Killing vector field $X$. Then $(M,g_{\mu\nu})$ will be
said to be: 
\begin{enumerate}
\item {\em Stationary}, if there exists an asymptotically Minkowskian
  coordinate system $\{y^\mu\}$ on 
  (perhaps a subset of)  $\Omega$, with $y^0$ --- a time coordinate,
  in which  $X=\partial/\partial y^0$.  When the orbits of $X$ are
complete we shall require that they are diffeomorphic to $\R$, and that
$\Sigma_R\equiv\{t=0,r(\vec x)\ge R\}$ intersects the orbits of $X$ only
once, at least for $R$ large enough.
\item {\em Axisymmetric}, if $X^\mu$ has complete periodic orbits.  
Moreover $X^\mu$ will be required to have an axis, that is, 
the set $\{p:X^\mu(p)=0\}\ne \emptyset$. 
\item {\em Stationary-rotating} (compare \cite{ChWald1}), if the
  matrix $\lam^\mu_\nu =\lim_{r\to\infty}{\partial_\nu
 X^\mu}$ is a rotation matrix, that is, $\lam^\mu_\nu$ has a
timelike eigenvector $a^\mu$, with zero eigenvalue\footnote{If 
$\lam^\mu_\nu$ has a
timelike eigenvector $a^\mu$, we can find a Lorentz frame so 
that $a^\mu=(a,0,0,0)$. In that frame $\lam^\mu_\nu$ satisfies
$\lam^0_\nu=\lam^\mu_0=0$, so that it generates 
space--rotations, if non--vanishing.}. Let $\phi_t[X]$ denote the flow
  of $X$. We shall moreover require that there exits $T>0$ such that
  $\phi_T[X](p)\in I^+(p)$ for $p$ in the exterior asymptotically
  flat 3-region $\Sext$.

\item {\em Stationary--axisymmetric}, if there exist on $M$ two
  {\em commuting} Killing vector fields $X_a$,  $a=1,2$, such that
$(M,g_{\mu\nu})$ 
is stationary with respect to $X_1$ and axisymmetric with respect to $X_2$,
\item {\em Spherically symmetric}, if, in an appropriate coordinate
system on $\Omega$,  $SO(3)$ acts on $M$ by
  rotations of the spheres $r={\rm const}$, $t={\rm const'}$
  in $\Omega$, at least for $t=0$ and $r$ large enough.
\item {\em Stationary--spherically symmetric}, if  $(M,g_{\mu\nu})$
is stationary   and spherically symmetric.
\end{enumerate}

We have the following:

\begin{Theorem}
  \label{T1} Under the conditions of Theorem \ref{T0}, let $G_0$
  denote the connected component of the group of all isometries of
  $(M,g_{\mu\nu})$.  If $G_0$ is non--trivial, then one of the
  following holds:
\begin{enumerate}
\item  $G_0=\R$, and  $(M,g_{\mu\nu})$ is either stationary, or
  stationary--rotating.
\item $G_0= U(1)$, and  $(M,g_{\mu\nu})$ is axisymmetric.
\item $G_0=\R\times U(1)$, and  $(M,g_{\mu\nu})$ is stationary--axisymmetric.
\item $G_0=SO(3)$, and  $(M,g_{\mu\nu})$ is spherically symmetric.
\item $G_0=\R\times SO(3)$, and  $(M,g_{\mu\nu})$ is
  stationary--spherically symmetric.   
\end{enumerate}
\end{Theorem}

We believe that the condition that $\Omega$ be a boost--type domain is
unnecessary. Recall, however, that this condition is reasonable for
vacuum space--times \cite{christodoulou:murchadha}, and one expects it
to be reasonable 
for a large class of couplings of matter fields to gravitation,
including   electro--vacuum space--times. We wish to point out that in
our proof that condition is needed  to exclude boost--type Killing
vectors, in Proposition \ref{PA.1} below, as well as 
to exclude causality violations
in the asymptotic region. We expect that it should be
possible to exclude the boost--type Killing vectors 
purely by an
initial data analysis, using the methods of \cite{ChBeig1}. If that
turns out to be the case, the only ``largeness requirements'' left on
$(M,g_{\mu\nu})$ would be the much weaker conditions\footnote{Those
global considerations of the proof of Theorem \ref{T1} which use the
structure of $\Omega$ can be carried  through under the condition
\eq{fcond}, provided that
the constants $C_1$ and $\hat C_1$ appearing there are replaced by some
appropriate larger 
constants. The reader should also note that these considerations are
unnecessary when $\Sigma_R$ is assumed to be achronal.}  needed in
Proposition \ref{PA.2} below. Let us also mention that in stationary
space--times with more than one Killing vector all the results below
can be proved directly by an analysis of initial data sets, so that no
``largeness'' 
conditions on $(M,g_{\mu\nu})$ need
to be imposed --- see \cite{ChBeigKIDs}.

Let us finally mention that the results here settle  in the positive
Conjecture 3.2 of \cite{Chnohair}, when  the supplementary
hypothesis of existence of at least two Killing vectors is made there.

We find it likely that there exist no electro--vacuum, asymptotically
flat space--times which have no black hole region, which are
stationary--rotating and for which $G_0=\R$. A similar statement
should be true for domains of outer communications of regular black
hole space--times. It would be of interest to prove this result. Let
us also point out that the Jacobi ellipsoids \cite{Chandra:ellipso}
provide a Newtonian example of solutions with a one dimensional group
of symmetries with a ``stationary--rotating'' behavior.

\section{Definitions, proofs.}
\label{proofs}

Let $W$ be a vector field, throughout we shall use the notation
$\phi_t[W]$ to denote the (perhaps defined only locally)  flow
generated by $W$.

Consider a subset $\Omega$ of $\R^4$ of the form
\begin{equation}
\Omega=\big\{(t,\vec x)\in {\R} \times {\R}^3: r((t,\vec x))\ge R, |t|\le
f(r(\vec x))\big\}\,, 
  \label{D.Omega}
\end{equation} 
for some constant $R\ge 0$ and some function $f(r)\ge 0, f\not\equiv
0$. We shall consider only {\em non--decreasing} functions $f$.  Here
and elsewhere, by a slight abuse of notation, we write 
$$
r((t,\vec x))=r(\vec x)=\sqrt{\sum_{i=1}^3 (x^i)^2}\,.
$$
 Let $\alpha$ be a
positive constant; $\Omega$ will be called 
{\em a boost--type domain} if $f(r)=\theta r+C$ for some constants
$\theta>0$ and $C\in\R$ ({\em cf.\/} also
\cite{christodoulou:murchadha}).

Let $\phi$ be a function defined on $\Omega$. For $\beta\in\R$ we
shall say that $\phi 
= \calo_k(r^{\beta})$ if $  \phi \in C^k(\Omega)$, and if there
exists a function $C(t)$ such that we have
$$
0\le i\le k \qquad |\partial_{\alpha_1}\cdots \partial_{\alpha_i}\phi
| \le C(t) (1+r)^{\beta-i}\ .
$$
We write $\calo(r^{\beta})$ for $\calo_0(r^{\beta})$.
We say that $\phi = o(r^{\beta})$  if $\lim_{r \ra \infty,
t=\mbox{\footnotesize  const}} r^{-\beta}\phi(t,x) = 0$. 
 A metric on $\Omega$ will be said to be asymptotically flat
 if there exist $\alpha>0$ and $k\in\N$  such that
\begin{equation}
\label{D1.new}
g_{\mu\nu} -\eta_{\mu\nu}= \calo_k(r^{-\alpha})\ ,
\end{equation}
 and if
there exists a function $C(t)$
such that 
\begin{eqnarray}
  & |g_{\mu\nu}| + |g^{\mu\nu}| 
\le C(t) \,, &
\label{D.1old}\\
& g_{00}\le -C (t)^{-1},\qquad g^{00}\le -C(t)^{-1} \,,&
\label{D.2old}\\
& \forall X^i\in {\R}^3 \quad g_{ij}X^iX^j\ge C (t)^{-1}\sum(X^i)^2 \,.&
\label{D.3old}
\end{eqnarray}
Here and throughout $\eta_{\mu\nu}$ is the Minkowski metric. 

Given a set $\Omega$ of the form \eq{D.Omega} with a metric satisfying
\eq{D1.new}--\eq{D.3old}, to every slice $\{t=\mbox{\rm
const}\}\subset\Omega$ one can  
associate in  a unique way the ADM four--momentum vector $p^\mu$ (see
\cite{ChErice,bartnik:mass}), provided that $k\ge 1$, $\alpha>1/2$, and
that the Einstein tensor satisfies the fall--off condition
\eq{Efalloff}. Those conditions also guarantee that $p^\mu$ will not depend 
upon which hypersurface $t=\mbox{\rm const}$ has been chosen. The ADM
four--momentum of $\Omega$ will be 
defined as the four--momentum of any of the hypersurface
$\{t={\rm const}\}\subset\Omega$.  

We note the following useful result:

\begin{Proposition}
  \label{PD.1}
Consider a metric $g_{\mu\nu}$ defined on a set $\Omega$ as in
\eq{D.Omega} (with a non--decreasing function $f$), and suppose that
$g_{\mu\nu}$ satisfies  \eq{D1.new}--\eq{D.3old} with $k\ge 2$.
Let $X^\mu$ be a Killing vector field defined on $\Omega$.
  Then there exist numbers $ \lam_{\mu\nu}= \lam_{[\mu\nu]} $
such that 
\begin{equation}
X^\mu-{\lam^\mu}_\nu x^\nu = \calo_k(r^{1-\alpha})\ ,
%|X^\mu-{\lam^\mu}_\nu x^\nu| + r |\partial_\sigma X^\mu -
%{\lam^\mu}_\sigma| + r^2 |\partial_\sigma \partial_\rho X^\mu | \le
%C'(t) r^{1-\alpha}\,,
  \label{D.4}
\end{equation}
with ${\lam^\mu}_\nu\equiv \eta^{\mu\alpha}\lam_{\alpha\nu}$.
If ${\lam_{\mu\nu}}=0$, then there exist numbers $A^\mu$ 
%and a function $C''(t)$ 
such that %on $M_R$ we have
\begin{equation}
X^\mu-A^\mu  = \calo_k(r^{-\alpha})\ .
%  |X^\mu-A^\mu | + r |\partial_\sigma X^\mu | + r^2 |\partial_\sigma
%  \partial_\rho X^\mu | \le C''(t) r^{-\alpha}\,.
  \label{D.5}
\end{equation}
If ${\lam_{\mu\nu}}=A^\mu=0$, then  $X^\mu \equiv 0$.
\end{Proposition}

\proof The result follows from Proposition 2.1 of \cite{ChBeig1},
applied to the slices $\{t=\mathrm{const}\}$, except for the estimates
on those partial derivatives of $X$ in which $\partial/\partial t$
factors occur.  Those estimates can be obtained from the estimates for
the space--derivatives of Proposition 2.1 of \cite{ChBeig1} and from
the equations \begin{equation} \label{I.0.new} \nabla_\mu\nabla_\nu
X_\alpha = {R^\lambda}_{\mu\nu\alpha}X_\lambda \ , \end{equation}
which are a well known consequence of the Killing equations.  \hfill
$\Box$

The proofs of Theorems \ref{T0} and \ref{T1} require several steps.
Let us start by showing that boost--type Killing vectors are possible
only if the ADM four--momentum is spacelike or vanishes:

\begin{Proposition}\label{PA.1} Let $g_{\mu\nu}$ be a twice
differentiable metric on a
  boost--type domain  $\Omega$, satisfying \eq{D1.new}--\eq{D.3old},
with $\alpha> 1/2$ and with $k\ge 
  2$. Suppose that the Einstein tensor $G_{\mu\nu}$ of $g_{\mu\nu}$
satisfies 
$$
G_{\mu\nu}= \calo(r^{-3-\epsilon}), \qquad \epsilon > 0\ .
$$  
 Let   $X^\mu$ be a  Killing vector field  on $\Omega $,
 set
  \be
  \label{lamdef}
  {\lam^\mu}_\nu \equiv \lim_{r\to\infty} \frac{\partial X^\mu}{\partial
   x^\nu}
   \ee
  (those limits exist by Proposition \ref{PD.1}).
   Then the ADM four--momentum $p^\mu$ of $\Omega$ satisfies   
\be
\label{rotation}
{\lam_\mu}^\nu p^\mu = 0\ .
\ee
\end{Proposition}
\proof If ${\lam^\mu}_\alpha=0$ there is nothing
to prove, suppose thus that ${\lam^\mu}_\alpha\ne 0$.
Let ${\Om^\mu}_\nu$ be a solution of the equation
$$
\frac{d{\Om^\mu}_\nu}{ds} = {\lam^\mu}_\alpha{\Om^\alpha}_\nu\ .$$
It follows from 
Proposition \ref{PD.1}  that the flow $\phi_t[W](p)$ is defined for
all $t\in [-\alpha, \alpha]$ and for all
$p\in\Sigma_{R_1}\equiv\{t=0,r(p)\ge R_1\}\subset\Omega$ for some constants
$\alpha$ and  $R_1$. By \cite[Theorem 1]{ChmassCMP}, in local
coordinates we have  
\begin{eqnarray*}
&\phi^\mu_t[W]={\Om^\mu}_\nu(t) x^\nu +\calo_{k}(r^{1-\alpha}) \ ,
& \\ &
\frac{\partial \phi^\mu_t[W]}{\partial x^\nu}={\Om^\mu}_\nu(t) +
\calo_{k-1}(r^{1-\alpha})\ .
&
\end{eqnarray*}
The error
terms above satisfy appropriate decay conditions so that the ADM
four--momentum 
$$
p_\mu(\phi_t[W](\Sigma_{R_1}))= \int_{\phi_t[W](\Sigma_{R_1})} 
U_\mu^{\alpha\beta} dS_{\alpha\beta}
$$ 
is finite and well--defined. Here 
$dS_{\mu\nu}=\iota_{\partial_\mu }\iota_{\partial_\nu}
dx^0\wedge \ldots \wedge dx^3$,
 $\iota_X$ denotes the inner product of a vector $X$ with a form, and 
 ({\em cf., e.g.}, \cite{ChmassCMP})
%Let $U^{\mu\nu}_\alpha$ be the Freud "superpotential":
$$
U^{\alpha\beta}_\mu = \delta^{[\alpha}_{\lambda}\delta^\beta_\nu
\delta^{\gamma]}_\mu \eta^{\lambda\rho} \eta_{\gamma\sigma}
{\partial_\rho g^{\nu\sigma} } \ . %dS_{\alpha\beta}\ .
$$
As is well known (see \cite{ChmassCMP} for a proof under
the current asymptotic conditions, {\em cf.\/} also
\cite{BOM:poincare,ashtekar:hansen}), under boosts the ADM
four--momentum transforms  like a four--vector, that is, 
\begin{equation}
  \label{I.1}
  p^\mu(\phi_t[W](\Sigma_{R_1}))= {\Om^\mu}_\nu(t)  p^\nu(\Sigma_{R_1})\ .
\end{equation}
On the other hand, the $\phi_t^\mu[W]$'s are isometries, so that
$$
g_{\alpha\beta}(\phi_t^\mu[W](x))
\frac{\partial\phi_t^\alpha [W]}{\partial x^\mu}(x)
\frac{\partial\phi_t^\beta [W]}{\partial x^\nu}(x)
= g_{\mu\nu}(x)\ ,
$$
which gives
\be
\label{**}
U^{\mu\nu}_\alpha (\phi_t^\mu[W](x)) {\Om ^\sigma}_\mu (t){\Om^\rho}_\nu (t)
= {\Om^\gamma}_\alpha(t)U_\gamma ^{\rho\sigma}(x) + \calo(r^{-1-2\alpha})
\ .
\ee
Eqs \eq{I.1} and \eq{**} give, for all $t$,
\be
\label{***}
{\Om ^\sigma}_\mu (t) p_\sigma = p_\mu\ ,
\ee
and \eq{rotation} follows by $t$--differentiation of eq.\ \eq{***}.
\hfill $\Box$

Suppose, now, that the ADM four--momentum $p^\mu$ of the hypersurface
$\{t=0\}$ is timelike. If $\Omega$ is large enough we can find a boost
transformation $\Lambda$ such that the hypersurface $\Lambda(\{t=0\})$
is asymptotically  orthogonal to
$p^\mu$.  It then follows by Proposition \ref{PA.1} that the matrix
$\lam$ defined in eq.~\eq{lamdef} has vanishing $0$-components in that
Lorentz frame, and therefore generates space rotations. We  need
to understand the structure of orbits of such Killing vectors. This is
analysed in the Proposition that follows:

\begin{Proposition}\label{PA.2} 
Let $g_{\mu\nu}$ be a metric on a set $\Omega$ as in eq.\
\eq{D.Omega}, and suppose  that  $g_{\mu\nu}$ satisfies the fall-off
condition \eq{D1.new} %--\eq{D.3} 
with $\alpha> 0$ and $k\ge 2$. 
Let $X^\mu$ be a  Killing vector field defined on $\Omega $, and suppose that 
\beq%a&
Z^\mu\partial_\mu \equiv 
X^\mu\partial_\mu  - {\omega^{i}}_j x^j \partial_i = o(r)\ , 
% & \label{L1.1} \\ &
\qquad 
\partial_\sigma Z ^\mu= o(1)\ .  %&
\label{L1.2}
\eeq%a
with $\omega^i{}_j$ --- a (non--trivial) antisymmetric matrix with
constant coefficients, normalized such that  $\omega^i{}_j
\omega^j{}_i=2(2\pi)^2$.  (It follows from Proposition
\ref{PD.1} that there exist  constants $C_1, \hat C_1$ such that $|X^0|\le
C_1r^{1-\alpha}+\hat C_1$ on $\{t=0\}\subset\Omega$.) Suppose that the
function $f$ in \eq{D.Omega} 
satisfies 
\beq
\label{fcond}
f(r)\ge  %2\pi 
C_2r^{1-\alpha}+\hat C_1\ ,
\eeq
where $C_2$ is any constant larger than $C_1$.
Let $\phi_s$ denote the flow of $X^\mu$. Then:
\begin{enumerate}
\item There exists $R_1 \geq R$ such that $\phi_s(p)$ is well defined
for $p \in\Sigma_{R_1}\equiv\{t=0, r\ge R_1\}\subset\Omega$ and  for $s
\in [0,1]$. For those values of $s$ we have  $\phi_s(\Sigma_{R_1})
\subset \Omega$. 
\item There exist constants $A^\mu $ such that, in local coordinates
on $\Omega$, for all $x^\mu\in\Sigma_{R_1}$  we have
\beq
\label{shifteq}
\phi^\mu_1= x^\mu + A^\mu + \calo_{k-1}(r^{-\alpha})\ .
\eeq
\item If $A^\mu=0$, then $\phi_1(p)=p$ for all $p$ for which
$\phi_1(p)$ is defined. 
\end{enumerate}
\end{Proposition}

{\bf Remark}: The hypothesis that $\lim_{r\to\infty}{\partial _i
  X^0}=0$, which is  made in \eq{L1.2}, is not needed for
points 2 and 3 above to hold, provided one assumes that the
conclusions of point 1 hold.

\proof Point 1 follows immediately from the asymptotic estimates
of Proposition \ref{PD.1} and the defining equations for $\phi^\mu_s$,
$$ \frac{d\phi^\mu_s }{ds}=X^\mu\circ\phi^\mu_s .$$ To prove point 2,
let $ {R^i}_j(s)$ be the solution of the equation 
\beqn
\frac{d { R^i}_j}{ds} = \omega^i{}_k { R^k}_j\ ,
\eeqn
with initial condition
${R^{i}}_{j} (0)=
\delta^i{}_j$, set ${R^{0}}_{0}(s) = 1$, ${R^0}_{i}(s) = 0$. 
We have the variation--of--constants formula
\beqn
\phi^\mu_s(x) = R^{\mu}_{\;\nu}(s) x^\nu + \int_0^s R^\mu{}_\nu(s-t)
Z^\nu(\phi_t(x))dt,
\eeqn
from which we obtain, in view of Proposition \ref{PD.1},
\beqa&
\frac{\partial \phi^\mu_{1}}{\partial x^\nu} - \delta^\mu{}_\nu
= \calo_{k-1}(r^{-\alpha}),& \label{E.o} \\&
\phi^\mu_{1} - x^\mu = \calo_k(r^{1-\alpha}). & \label{E.o.1}
\eeqa
Set $y^\mu(x) = \phi^\mu_{1}(x)$. As $y^\mu(x^\nu)$ is an isometry, we
have the equations
\beq
\frac{\partial^2 y^\alpha}{\partial x^\mu \partial x^\nu} =
\Gamma^\sigma_{\mu\nu}(x) \frac{\partial y^\alpha}{\partial x^\sigma}
- \Gamma^\alpha_{\beta\gamma}(y(x)) \frac{\partial y^\rho}
{\partial x^\mu} \frac{\partial y^\gamma}{\partial x^\nu}.
\label{E.1}
\eeq
{}From \eq{E.o}--\eq{E.o.1}  we obtain
\beqa
\frac{\partial^2 (y^\alpha-x^\alpha)}{\partial x^\mu \partial x^\nu}  & = & 
\Gamma^\alpha_{\mu\nu}(x) 
- \Gamma^\alpha_{\mu\nu}(y(x))+ \calo_{k-1}(r^{-1-2\alpha})
\nonumber
\\
& = & 
(y^\rho(x)-x^\rho)\int_0^1\partial_\rho \Gamma^\alpha_{\mu\nu}(tx
+ (1-t)y(x))dt+ \calo_{k-1}(r^{-1-2\alpha})
\nonumber
\\
 & = &  \calo_{k-2}(r^{-1-2\alpha}) \ .
\label{Neq.1}
\eeqa
%(To obtain the last inequality we have used the mean value theorem and
%the fall--off behaviour of the second derivatives of the metric.)
We can integrate this inequality in $r$ 
to obtain 
$$
\frac{\partial(y^\alpha - x^\alpha)}{\partial x^\mu} = 
\calo_{k-1}(r^{-2\alpha}) \ .
$$
If $2\alpha > 1$, the Lemma of 
the Appendix A of \cite{ChmassCMP} shows  that the limits
$\lim_{r\to\infty,t=0} (y^\alpha-x^\alpha) = A^\alpha$ exist and
we get
$$
y^\alpha - x^\alpha = A^\alpha +\calo_{k}(r^{1-2\alpha}) \ .
$$
Otherwise, decreasing $\alpha$ slightly if necessary, we may assume
that  $2\alpha < 1$, in which case we simply obtain
$$
y^\alpha - x^\alpha = \calo_{k}(r^{1-2\alpha}) \ .
$$
If the last case occurs we can  repeat this argument 
$\ell-1$ times to
obtain $ \calo(r^{-1-(\ell+1)\alpha})$ 
at the right--hand--side of \eq{Neq.1} until $-1-(\ell+1)\alpha< -2$;
at the last iteration we shall thus  obtain  
$ \calo(r^{-2-\epsilon})$ there, 
with some $\epsilon >0$. We can again use the Lemma of 
the Appendix A of \cite{ChmassCMP} to conclude that the limits
$\lim_{r\to\infty,t=0} (y^\alpha-x^\alpha) = A^\alpha$ exist. An
iterative argument similar to the one above applied to \eq{Neq.1}
gives then 
\beq
\xi^\alpha\equiv  y^\alpha- x^\alpha - A^\alpha
=\calo_{k}(r^{-\alpha})\ ,\label{Neq.3}
\eeq
which establishes point 2.

Suppose finally that $A^\mu$ vanishes. 
Eq.\ (\ref{E.1}) implies an inequality of the form
\beq
\left| \frac{\partial^2(y^\alpha - x^\alpha)}{\partial x^\mu \partial
x ^\nu}\right| \leq C(|\partial \Gamma| |y-x| + |\Gamma|
|\partial(y-x)|), \label{E.2}
\eeq
for some constant $C$. A standard bootstrap argument using (\ref{E.2}), 
(\ref{E.o}) and (\ref{E.o.1}) shows that for all $\sigma \geq 0$ we have
\beq
\lim_{r \ra \infty} [ r^{\sigma}|y-x| + r^{\sigma}|\partial(y-x)|] = 0.
\label{E.3}
\eeq
Define
\beq
F = r^{\beta-2} |y-x|^2 + r^\beta|\partial(y-x)|^2.
\eeq
Choosing $\beta$ large enough one finds from (\ref{E.2}) that
\beq
\frac{\partial F}{\partial r} \geq 0.
\eeq
This implies
\beq
R_2 \leq r \leq r_1 \Ra F(r_1) \geq F(r) \geq 0.
\eeq
Passing with $r_1 \ra \infty$  from (\ref{E.3}) we obtain
$\phi_1(x)=x$ for $x \in\Sigma_{R_1}$.  $\phi_1$ is therefore an isometry
which reduces to an identity on a spacelike hypersurface, and point 3
follows from \cite[Lemma 2.1.1]{SCC}. 
\hfill\ $\Box$

We are ready now to pass to the proof of Theorem \ref{T0}:

{\bf Proof of Theorem \ref{T0}}:
Let $y^\alpha(x^\beta)$ be defined as in the proof of Proposition
\ref{PA.2}, as it is an isometry we have the equation:
\beq
g_{\mu\nu}(y(x))\frac{\partial y^\mu}{\partial  x^\alpha}
\frac{\partial y^\nu}{\partial  x^\beta}= 
g_{\alpha\beta}(x)\ .
\label{Neq.4}
\eeq
Set $\xi_\alpha=\eta_{\alpha\beta}\xi^\beta$, where
$\eta_{\alpha\beta}={\rm diag}(-1,1,1,1)$, with $\xi$ defined by eq.
\eq{Neq.3}. Eqs.\ \eq{Neq.3}   and
\eq{Neq.4} together with the asymptotic form of the metric, eq.\
\eq{D1.new}, give 
\beq
\frac{\partial \xi_\alpha}{\partial  x^\beta}+
\frac{\partial \xi_\beta}{\partial  x^\alpha}+
g_{\alpha\beta}(x^\sigma+A^\sigma+\xi^\sigma)-g_{\alpha\beta}(x^\sigma)
 =\calo_{k-1}(r^{-1-2\alpha})\ .
\label{Neq.5}
\eeq
Suppose first that $A^\sigma\not \equiv 0$; 
%by the mean value theorem
%there exists $\theta(x)\in[0,1]$ such that
we have
\begin{eqnarray*}
& g_{\alpha\beta}(x^\sigma+A^\sigma+\xi^\sigma)-g_{\alpha\beta}(x^\sigma)
 = \qquad \qquad \qquad  \qquad \qquad \qquad \qquad \qquad \qquad \qquad \qquad \qquad& \\
& %\qquad \qquad \qquad
 =
%\frac{\partial
%g_{\alpha\beta}}{\partial
%x^\rho}\Big|_{(x^\sigma+\theta(x^\sigma)(A^\sigma+\xi^\sigma))} 
% (A^\rho+\xi^\rho) 
\frac{\partial g_{\alpha\beta}}{\partial x^\rho}{(x^\sigma)} 
 A^\rho+\int_0^1\left(\frac{\partial g_{\alpha\beta}}{\partial
x^\rho}{(x^\sigma+s(A^\sigma+\xi^\sigma))} (A^\rho+\xi^\rho))-
\frac{\partial g_{\alpha\beta}}{\partial x^\rho}{(x^\sigma)}A^\rho\right)ds &
\\
&  = 
\frac{\partial g_{\alpha\beta}}{\partial x^\rho}%\Big|_
{(x^\sigma)} 
 A^\rho + \calo(r^{-1-2\alpha})\ .
\qquad \qquad \qquad  \qquad \qquad \qquad \qquad %\qquad 
\label{Neq.6.0}
\end{eqnarray*}
A similar calculation for the derivatives of $ g_{\alpha\beta}$ gives
\beq
\label{Neq.6.1}
g_{\alpha\beta}(x^\sigma+A^\sigma+\xi^\sigma)-g_{\alpha\beta}(x^\sigma)
 =\frac{\partial g_{\alpha\beta}}{\partial x^\rho}%\Big|_
{(x^\sigma)} 
 A^\rho +\calo_{k-2}(r^{-1-2\alpha})\ .
\eeq
In a neighbourhood of 
$\Sigma_{R_1}$ define a vector field $Y^\mu$ by
$$
Y^\mu=\xi^\mu+A^\mu\ .
$$
It follows from \eq{Neq.5}--\eq{Neq.6.1} that $Y^\mu$ satisfies the
equation 
$$
\nabla_\mu Y_\nu + \nabla_\nu Y_\mu  = \calo_{k-2}(r^{-1-2\alpha})
\ .
$$
By hypothesis we have $k\ge 3$ and $2\alpha >1$,
we can thus use \cite[Proposition 3.1]{ChBeig1} 
%(with $N= -Y^\mu n_\mu$,
%where $n^\mu$ is the future directed normal to the slices
%$t=\mathrm{const}$) 
to conclude that $A^\mu$ must be proportional to
$p^\mu$. The remaining claims follow directly by Proposition
\ref{PA.2}. \hfill $\Box$

To prove Theorem \ref{T1} we shall need two  auxiliary results:

\begin{Proposition}\label{PR} 
Under the hypotheses of Prop.\ \ref{PD.1}, let $W$ be a non--trivial
Killing vector  field defined on %a set 
$\Omega$. 
% \subset M_R$ such that $\Omega \supset\Sigma_R$. Suppose that on
% $\Omega$ the decay conditions 
%\beq
%g_{ij} - \delta_{ij} = O_2(r^{-\ep}), \qquad
%K_{ij} = O_1(r^{-1-\ep}), \qquad
%T_{\mu\nu} = \calo(r^{-2-\ep}), \qquad \ep > 0
%\eeq
%hold. 
Suppose that there exists $R_1$ such that for $p\in\Sigma_{R_1}$ the
orbits $\phi_s[W](p)$ are defined for $s\in[0,1]$, with
$\phi_1[W](p)=p$. Assume moreover that there exists a non--vanishing
antisymmetric 
matrix with constant coefficients ${\omega^i}_j$ such that
$W^\mu\partial_\mu-{\omega^i}_jx^j\partial_i = o(r)$. 
%If the orbits of $W$ are periodic on $\Omega$, t
Then the set $\{p:W(p) = 0\}$ is not empty.
\end{Proposition}

\paragraph{Remark:}
%\begin{enumerate}
%\item The set $\Omega$ here could coincide with $\Sigma_R$ or perhaps with
%$M_R$. We wish to stress that no other conditions than $\Omega \supset
%\Sigma_R$ are imposed.
%\item For the result under consideration we are {\em not} assuming the
%existence of a time--like Killing vector on $\Omega$.
%\item 
The following half--converse to the Proposition \ref{PR} is well known:
Let W be a Killing vector field on a Lorentzian manifold $M$ and suppose that
$W(p) = 0$. If there exists a neighborhood $\cal O$ of $p$ such that
$W$ is nowhere time--like on $\cal O$, then there exists $T > 0$ such
that all orbits which are defined for $t \geq T$ are periodic.
%\end{enumerate}

\paragraph{Proof:} 
%Suppose that the orbits of $W$ are periodic on
%$\Omega$. As $\Omega$ is strongly causal (being a subset of the strictly
%causal manifold $M_R$), it follows by simple asymptotic considerations
%based on Prop. \ref{PD.1} that
%\beq
%\lim_{r \ra \infty} W^0{}_{,\nu} = \lim_{r \ra \infty} 
%W^\nu{}_{,0} = 0,
%\eeq
%\beq
%\lim_{r \ra \infty} W^i{}_{,j} = \omega_{ij} \not\equiv 0,
%\eeq
%where $\omega_{ij}$ is a rotation matrix. 
%For $s \in {\R}$ l
Let $\phi_s$ denote the flow of $W$ on $\Omega$, 
and for $p\in\Sigma_{R_1}$ define
\beqa
\bar t(p ) &=& \int_0^1 t \circ \phi_s(p ) ds ,\label{tint} \\
\bar r(p ) &=& \int_0^1 r \circ \phi_s(p ) ds.\label{rint}
\eeqa
%Here $T$ is the smallest (non--zero) period of $\phi_t$. 
Note that ${(\phi_s)}_*$ asymptotes to the matrix ${R^\mu}_\nu(s)$
defined in the proof of Prop. \ref{PA.2}, which gives 
$$ %\beq
\nabla \bar r = \int_0^1 {(\phi_s)}_*(\nabla r) \circ \phi_s(p ) ds
\approx \nabla r + \calo(r^{-\alpha}).
$$ %\eeq
Similarly
$$ %\beq
\nabla \bar t \approx \nabla t + \calo(r^{-\alpha}).
$$ %\eeq
This shows that for $R$ large enough the sets
$S_{R,T} = \{ p: \bar r(p ) = R,\  \bar t(p ) = T\}$ are
differentiable spheres. Moreover
\beq
\bar r \circ \phi_s = \bar r, \qquad
\bar t \circ \phi_s = \bar t,
\eeq
so that $W$ is tangent to $S_{R,T}$. As every continuous vector
field tangent to a two--dimensional sphere has fixed points, the
result follows. \hfill\ $\Box$

{\bf Proof of Theorem \ref{T1}:} Let $\Lie$ denote the Lie algebra of
$G_0$, to any element $h$  of $\Lie$ there is associated a unique Killing
vector field $X^\mu(h)$, the orbit of which is complete.

Suppose first  that $\Lie$
is 1--dimensional. If the constant $a$ of  Theorem \ref{T0}
vanishes,  $(M,g_{\mu\nu})$ is axisymmetric by part 3 of  Theorem
\ref{T0} and 
by Proposition \ref{PR}. If $a$ 
does not vanish there are two cases to analyse.
Consider first the case in which $\partial_\mu X^\nu\not\rightarrow 0$
as $r\rightarrow \infty$. Let us perform a Lorentz transformation so that
the new hypersurface $t=0$, still denoted by $\Sigma_R$, is
asymptotically normal to $p^\mu$.  
By Proposition \ref{PA.1} we must have
$\lim_{r\to\infty}\partial_i X^0 =\lim_{r\to\infty}\partial_0
X^i=0$,  hence Proposition \ref{PA.2} applies. As $M$
contains a boost--type domain for any $T$ we can choose $p\in
\Sigma_{R_1} $, with $r(p)$ large enough, so that $\phi_s[X](p)$ is
defined for all $s\in[0,T]$, with $\phi_s[X](p) \ne p$ by
\eq{shifteq}. This shows that $G_0$ cannot be $U(1)$, hence 
$G_0=\R$, and $(M,g_{\mu\nu})$ is stationary--rotating as claimed. 
The second case to consider is, by Proposition \ref{PD.1}, that in
which $ X^\mu\rightarrow ap^\mu=A^\mu$ as $r\rightarrow \infty$ in
$\Omega$. We want to show that $\Sigma_{R}$ is a global
cross--section for $\phi_s[X]$, at least for $R$ large enough. To do
that, note that timelikeness of $A^\mu$ implies that we can choose
$R_2$ large enough so that $X^\mu$ is transverse to
$\Sigma_{R_2}$. Let $(g_{ij},K_{ij})$ be the induced metric and the
extrinsic curvature of $\Sigma_{R_2}$, and let $(\hat M, \hat
g_{\mu\nu})$ be the Killing development of
$(\Sigma_{R_2},g_{ij},K_{ij})$ constructed using the Killing vector
field  $ X^\mu $, see Section 2 of
\cite{ChBeig1} for details. Define $\Psi:\hat
M\rightarrow M_{R_2}\equiv\cup_{t\in\R} \phi_t[X](\Sigma_{R_2})$ by
$\Psi(t,\vec x)= \phi_t[X](0,\vec x)$. Then $\Psi$ is a local
isometry between $\hat M$ and $
M_{R_2}$. $\Psi$ is surjective by construction, and there exists a
boost--type domain $\hat \Omega$ in $\hat M$ such that $\Psi|_{\hat
\Omega}$ is a diffeomorphism between  $\hat \Omega$ and $ \Omega$.

Suppose that $\Psi$ is not injective, let us first show that this is
equivalent to the statement that $\Psi^{-1}(\Sigma_{R_2})$ is not
connected. Indeed, let $p=(t,\vec x)$ and $q = (\tau,\vec y)$ be such
that $\Psi(p)=\Psi(q)$, then $\phi_{-t}(\Psi(p))=\phi_{-t}(\Psi(q))$ so
that
$\Psi((0,\vec x))=\Psi((\tau-t,\vec y))$, which leads to $
(\tau-t,\vec y)\in \Psi^{-1}(\Sigma_{R_2})$. 

Consider any connected component $\hat \Sigma$ of
$\Psi^{-1}(\Sigma_{R_2})$, as $\Psi$ is a local isometry 
$\hat \Sigma$ is an asymptotically flat hypersurface in $\hat M$.
By \cite[Lemma 1 and Theorem 1]{ChmassCMP}, we have
$$
\hat \Sigma = \{ t=h(\vec x), \quad \vec x \in {\cal U}\in \R^3\}\ ,
$$
where $\cal U$ contains $\R^3\setminus B(R_3)$ for some $R_3\ge
R_2$. Morever there exists a Lorentz matrix ${\Lambda^\mu}_\nu$ such
that 
$$
h(\vec x) = {\Lambda^0}_iX^i + O(r^{1-\alpha})\ .
$$
Note that the unit normal to $\hat \Sigma$ approaches, as
$r\to\infty$, the Killing vector $X$, hence
$$
{\Lambda^\mu}_\nu X^\nu = X^\mu \quad \Rightarrow \quad 
{\Lambda^0}_i= {\Lambda^i}_0= 0\ .
$$
It follows that $h(\vec x)=O(r^{1-\alpha})$, so that $\Psi((h(\vec
x),\vec x))\in \Omega$ for %$r(\vec x)$ large enough, 
$r(\vec x)\ge R_4$ for some constant $R_4\ge R_3$. 

Consider a point $q\in \Sigma_{R_4}$, then there exists a point
$(0,\vec x)$ such that $\Psi(0,\vec x)=q$ and a point $(h(\vec y),\vec
y)\in \hat\Sigma$ such that $\Psi(h(\vec y),\vec y))=q$. This, however,
contradicts that fact that $\Psi|_{\hat\Omega}$ is a diffeomorphism
between the boost-type domain $\hat\Omega$ and $\Omega$. We conclude
that
%As no %
% points $x^\mu$ and $y^\mu$ in $\Omega$ such that
%$x^0-y^0=\calo(r^{1-\alpha})$ are identified 
% we conclude that $H=\{id\}$, so that 
$\psi$ is injective. It follows that $\psi$ is a bijection, which
implies that all  the orbits through $p\in \Sigma_{R_2}$ are
diffeomorphic to $\R$, and that they intersect $\Sigma_{R_2}$ only once.

Suppose next that $\Lie$ is two--dimensional. Then there exist on $M$
two linearly independent Killing vectors $X^\mu_a$, $a=1,2$. Propositions
\ref{PA.1} and \ref{PA.2} lead to the following three possibilites: 

i) There exist constants $B^\mu_a$, $a=1,2$ such that
$X^\mu_a-B^\mu_a=o(1)$. By %Theorem \ref{T0} and 
\cite[Proposition 3.1]{ChBeig1} we have $ B^\mu_a=a_ap^\mu$ for some
constants $a_a$. 
It follows that there exist constants $(\alpha,\beta)\not = (0,0)$
such that $\alpha X_1^\mu + \beta X^\mu_2 =o(1)$. Proposition
\ref{PD.1} implies that  $\alpha X_1^\mu + \beta X^\mu_2 =0$, which
contradicts the hypothesis $\dim \Lie = 2$, therefore this case cannot
occur. 

ii) There exist constants $B^\mu$ and ${\omega^i}_{j}=-{\omega^j}_{i}$
such that 
\beq
\label{Neq.7}
X^\mu_1-B^\mu=o(1), \qquad X^\mu_2\partial_\mu-
{\omega^i}_{j}x^i\partial_j = o(r)\ .
\eeq
Consider the commutator $[X_1,X_2]$. The estimates on the derivatives
of $X_a^\mu$ of  Proposition \ref{PD.1} give
$[X_1,X_2]^0 = o(1)$, $[X_1,X_2]^i = o(r)$, so that by
Prop. \ref{PD.1} the commutator $[X_1,X_2]$ 
either vanishes, or asymptotes a constant vector with
vanishing time--component, hence spacelike. The latter case cannot
occur in view  of \cite[Proposition
3.1]{ChBeig1}, hence $[X_1,X_2]=0$. It follows that $\phi_t[X_2+\alpha
X_1]= \phi_t[X_2]\circ\phi_t[\alpha X_1]$.
%, whenever all the quantities
%occurring there are defined. 
Let $ap^\mu$ be the vector  given by
Theorem \ref{T0} for the vector field $X_2^\mu$. In local coordinates
we obtain 
$$\phi_1^\mu[X_2+\alpha
X_1]= x^\mu + ap^\mu +\alpha B^\mu +\calo(r^{-\alpha})\ .
$$
By  \cite[Proposition 3.1]{ChBeig1} we have $B^\mu \sim p^\mu$, so
that we can choose $\alpha$ so that $ \phi_1^\mu[X_2+\alpha
X_1]= x^\mu +\calo(r^{-\alpha})$. By point 3 of Theorem \ref{T0} we obtain
$\phi_1[X_2+\alpha X_1](p)=p$, hence all orbits of $X_2^\mu+\alpha
X_1^\mu$ are periodic with period $1$. As $p^\mu$ is time--like, the
orbits of $X_1^\mu$ must be time--like in the asymptotic region. 
%Since $\{t=0\}$ is achronal, 
As before, those orbits cannot be periodic because the coordinates on
$\Omega$ cover a boost--type region, hence they must be diffeomorphic
to $\R$. As 
$[X_1,X_2]=0$, we obtain that $G_0$ is the direct product $\R\times
U(1)$.

iii) For $\dim \Lie=2$ the last case left to consider is that when there
exist non--zero constants $\omega_{ij}^a$,
 $a=1,2$, such that $X_a^\mu\partial_\mu-\omega^a_{ij}x^i\partial_j =
o(r)$. Suppose that the antisymmetric matrices  $\omega_{ij}^a$ do not
commute, then by well known properties of $so(3)$ the matrices
$\omega_{ij}^a$ together with the matrix
$\omega_{ij}^1\omega_{jk}^2-\omega_{ij}^2\omega_{jk}^1$ are linearly
independent. It follows that  $[X_1,X_2]$ is a Killing vector linearly
independent of $X_1$ and $X_2$ near infinity, whence everywhere in
$\Omega$''.
 It is well known that the orbits of
$[X_1,X_2]$ are complete when those of $X_1$ and $X_2$ are 
\cite[Theorem 3.4]{kobayashi:nomizu}, which
implies that $G_0$ is at least three--dimensional,   which contradicts
$\dim \Lie=2$. If the matrices $\omega_{ij}^a$ commute they are
linearly dependent. Thus there 
exist constants $(\alpha,\beta)\not = (0,0)$ such that $\alpha X_1^\mu +
\beta X^\mu_2=o(r)$. By
Proposition \ref{PD.1} the Killing vector field $\alpha X_1^\mu +
\beta X^\mu_2$ is a translational Killing vector,  and the case here
is reduced to point ii) above.

Let us turn now to the case of a three dimensional Lie algebra
$\Lie$. An analysis similar to the above shows that this can only be
the case if three Killing vector fields $X_i^\mu$, $i=1,2,3$, on $M$
can be chosen so that
$X_i^\mu\partial_\mu-\epsilon_{ijk}x^j\partial_k=o(r)$. Moreover we
must have $[X_i,X_j]=\epsilon_{ijk}X_k$. Then $\Lie$ is the Lie
algebra of $SO(3)$, so that $G_0=SO(3)$, or its covering group
$Spin(3)=SU(2)$ \cite[p.\ 117, Problem 7]{Kirillov}.
 Integrating over the group as in the
proof of Proposition \ref{PR} (the integral $\int_0^1$ in eqs.\
\eq{tint}--\eq{rint} should be
replaced by an integral over the group $G_0$ with respect to the Haar
measure) one can pass to a new coordinate system,
defined perhaps only on a subset of $\Omega$,  such that the spheres
$t={\rm const}$,  $r={\rm const'}$ are invariant under
$G_0$. 
%The  asymptotic behavior of the $X^\mu_i$'s shows that
$G_0$ must be $SO(3)$, as $SO(3)$ is\footnote{This can be seen as
  follows: Any isometry is uniquely determined by its action at one
  point of the tangent bundle. Since $SO(3)$ acts transitively on
  $TS^2$, no larger groups can act effectively there.} the largest
group acting effectively on $S^2$. The proof of point 5) is left to
the reader. \hfill $\Box$

\section{Concluding remarks} 
\label{conclusions}

 Theorem \ref{T0} leaves open the intriguing possibility of a
space--time which has {\em only one} Killing vector which, roughly speaking,
behaves as a spacelike rotation accompanied by a time--like
translation. We conjecture that this is not possible 
%(unless the space--time is stationary), 
when the Einstein tensor $G_{\mu\nu}$
falls--off at a sufficiently fast rate, when global regularity
conditions are imposed 
and when positivity conditions on $G_{\mu\nu}$ are imposed.

One would like to go beyond the classification of groups given here,
and consider the  whole group of isometries $G$, not only the connected
component of the identity thereof $G_0$. Recall, {\em e.g.}, that a discrete
group of conformal isometries  acts on the critical space--times
which arise in the context of the Choptuik effect
\cite{Choptuik,Gundlach}. Let us first consider the case of
time--periodic space--times. Clearly such  space--times
exist when no field equations or energy inequalities hold, so
that the classification question becomes interesting only when some
field equations or energy--inequalities are imposed. In the vacuum
case some stationarity results 
have been obtained for spatially compact space--times by Galloway
\cite{galloway-splitting}. In the asymptotically flat context 
non--existence of periodic non--stationary vacuum solutions with an
analytic Scri has been established by  Papapetrou \cite{Papapetrou:periodic},
{\em cf.\/} also  Gibbons and Stewart
\cite{GibbonsStewart:periodic}. The
hypothesis of analyticity of Scri is, however, difficult to justify;
moreover the example of 
boost--rotation symmetric space--times shows that the condition of
asymptotic flatness in light--like directions might lead to
essentially different behaviour, as compared to that which arises in
the context of  asymptotic flatness in space--like directions. 
One expects that  non--stationary time--periodic vacuum
space--times do not exist, but no satisfactory analysis of that possibility
seems to have been done so far.

Another set of discrete isometries that might arise is that of
discrete subgroups of the rotation group, time--reflections,
space--reflections, etc. In those cases $G/G_0$ is compact.
It is easy to construct initial data $(g_{ij},K_{ij})$ on a compact or
asymptotically 
flat manifold $\Sigma$ which are invariant under a discrete isometry
group, in such a way that the group $H$ of all isometries of $g_{ij}$
which preserve $K_{ij}$ is {\em not} connected. By \cite[Theorem
2.1.4]{SCC} the group  $H$ will act by isometries on the maximal
globally hyperbolic development $(M,g_{\mu\nu})$ of
$(\Sigma,g_{ij},K_{ij})$, and it is rather clear that in generic such
situations the  groups $G$ of all isometries of $(M,g_{\mu\nu})$
will coincide with $H$. In this way one obtains space--times in
which $G/G_0$ is compact. It is tempting to conjecture that for, say 
vacuum, globally hyperbolic space--times with a compact or
asymptotically flat, appropriately regular, Cauchy surface, the
quotient $G/G_0$  will be a finite set. The proof of such a result would
imply non--existence of non--stationary time--periodic space--times,
in this class of space--times.

 {\bf  Acknowledgments}
P.T.C. is grateful to the E. Schr\"odinger Institute and to the
Relativity Group in Vienna for hospitality during part of work on this
paper. We are grateful to A. Fischer and A. Polombo for useful comments.

\bibliographystyle{amsplain}       
%\bibliographystyle{ieeetr} %orders references in the order of
                           %appearence in the text, and cuts the
                                        %second initial of the author's name     
%\bibliographystyle{siam}  % simply awful, takes authors initials in
                           %the quotation mark   
%\bibliographystyle{plain} 
%\bibliographystyle{apalike} % authors written IN the reference bracket!
%\bibliographystyle{unsrt}  % looks the same as plain to me?    
\bibliography{$HOME/prace/references/hip_bib,%
$HOME/prace/references/reffile,%
$HOME/prace/references/newbiblio}
%$HOME/prace/references/addon}
\end{document}